\documentclass[12pt]{article}
\usepackage{graphicx}

\begin{document}

\title{\textbf{Possible Way to Synthesize Superheavy Element Z=117 \footnote{Supported by
the National Natural Science Foundation of China under Grant No.
10475100, 10505016, the Knowledge Innovation Project of the Chinese
Academy of Sciences under Grant No KJCX2-SW-N17, KJCX-SYW-N2, and
Major state basic research development program 2007CB815000.}}}

\author{FENG Zhao-Qing$^{1,3} \footnote {Email: fengzhq@impcas.ac.cn}$, JIN Gen-Ming$^{1,2}$,
HUANG Ming-Hui$^{1}$, \\[0pt] GAN Zai-Guo$^{1}$, WANG Nan$^{4}$, LI Jun-Qing$^{1,2}$}
\date{}
\maketitle

\begin{center}
$^{1}${\small \emph{Institute of Modern Physics, Chinese Academy of Sciences,\\[0pt]
Lanzhou 730000, China}}\\[0pt]
$^{2}${\small \emph{Center of Theoretical Nuclear Physics,
National Laboratory of Heavy Ion Accelerator of Lanzhou,\\[0pt]
Lanzhou 730000, China}}\\[0pt]
$^{3}${\small \emph{Gesellschaft f\"{u}r Schwerionenforschung mbH
(GSI) D-64291 Darmstadt, Germany}}\\[0pt]
$^{4}${\small \emph{School of Physics, Shenzhen University, Shenzhen 518060, China}}\\[0pt]
\end{center}

\begin{abstract}
Within the framework of the dinuclear system model, the production
of superheavy element Z=117 in possible projectile-target
combinations is analyzed systematically. The calculated results show
that the production cross sections are strongly dependent on the
reaction systems. Optimal combinations, corresponding excitation
energies and evaporation channels are proposed in this letter, such
as the isotopes $^{248,249}Bk$ in $^{48}Ca$ induced reactions in 3n
evaporation channels and the reactions $^{45}Sc+^{246,248}Cm$ in 3n
and 4n channels, and the system $^{51}V+^{244}Pu$ in 3n channel.
\end{abstract}


\bigskip

The synthesis of superheavy elements has obtained much progress
experimentally using cold fusion reactions with double magic nucleus
$^{208}Pb$ or nearly magic nucleus $^{209}Bi$ as targets and hot
fusion reactions with double magic nucleus $^{48}Ca$ bombarding
actinide nuclei in fusion-evaporation reaction
mechanism.$^{\cite{Hof,Og1}}$ Especially in recent years, superheavy
elements Z=113-116, 118 have been synthesized successfully in
laboratories.$^{\cite{Og1,Mor}}$ Of course, further confirmations on
these new elements have to been done. Various theoretical models
have also been proposed to describe the formation of superheavy
nucleus.$^{\cite{Ada,Ari,Zag,Liu,Li1,Fe1}}$ Among these models, the
dinuclear system (DNS) model can describe very well a series of
available experimental data.$^{\cite{Ada,Li1}}$ We have further
developed the DNS model by taking the diffusion process to be
coupled with the relative motion, and by introducing barrier
distribution function method in the capture of two colliding nuclei
and in the fusion process.$^{\cite{Fe2}}$ The decay of the formed
DNS (quasi-fission) and the fission of heavy fragment are also
considered in the model. For all of reaction systems including cold
fusion, $^{48}Ca$ induced reactions and other projectile-target
combinations etc, we can use the same fixed parameters to reproduce
experimental data.

In the DNS concept, the compound nucleus is formed by nucleon
transfer from the light nucleus to the heavy one by overcoming the
inner fusion barrier. The evaporation residue cross section can be
written as a sum over all partial waves J at the centre-of-mass
energy $E_{c.m.}$,
\begin{equation}
\sigma_{ER}(E_{c.m.})=\frac{\pi \hbar^{2}}{2\mu
E_{c.m.}}\sum_{J=0}^{J_{max}}(2J+1)
T(E_{c.m.},J)P_{CN}(E_{c.m.},J)W_{sur}(E_{c.m.},J).
\end{equation}
Here $T(E_{c.m.},J)$ is the probability of two colliding nuclei
overcoming the potential barrier in the entrance channel to form the
DNS. $P_{CN}(E_{c.m.},J)$ is the probability that the system will
evolve from a touching configuration into the formation of compound
nucleus against quasi-fission of the DNS and fission of the heavy
fragment. The last term is the survival probability of the formed
compound nucleus, in which the fissile compound nucleus will decay
by emitting $\gamma$ rays, neutrons and charged particles etc. In
the calculation we take the maximal angular momentum as $J_{max}=30$
since the fission barrier of the heavy nucleus disappears at very
high spin.$^{\cite{Rei}}$

The same as in the nucleon collectivization model,$^{\cite{Zag}}$
the transmission probability is calculated by introducing a barrier
distribution function, which can reproduce very well available
experimental capture cross sections.$^{\cite{Zag,Fe2}}$ The fusion
probability is obtained by solving Master Equation numerically in
the potential energy surface of the DNS. The time evolution of the
distribution function $P(A_{1},\varepsilon^{\ast}_{1},t)$ for
fragment 1 with mass number $A_{1}$ and local excitation energy
$\varepsilon^{\ast}_{1}$ is described by the following Master
Equation,
\begin{eqnarray}
\frac{d P(A_{1},\varepsilon^{\ast}_{1},t)}{dt}=\sum_{A_{1}^{\prime
}}W_{A_{1},A_{1}^{\prime}}(t)\left[
d_{A_{1}}P(A_{1}^{\prime},\varepsilon^{\ast\prime}_{1},t)-d_{A_{1}^{\prime
}}P(A_{1},\varepsilon^{\ast}_{1},t)\right]-
\nonumber \\
\left[\Lambda^{qf}(\Theta(t))+\Lambda^{fis}(\Theta(t))
\right]P(A_{1},\varepsilon^{\ast}_{1},t).
\end{eqnarray}
Here $W_{A_{1},A_{1}^{\prime}}$ is the mean transition probability
from the channel $(A_{1},\varepsilon^{\ast}_{1})$ to
$(A_{1}^{\prime},\varepsilon^{\ast\prime}_{1})$, while $d_{A_{1}}$
denotes the microscopic dimension corresponding to the macroscopic
state $(A_{1},\varepsilon^{\ast}_{1})$. The sum is related to
$W_{A_{1},A_{1}^{\prime}}$ and taken as $A_{1}^{\prime }=A_{1}\pm 1$
for one nucleon transfer. The local excitation energy
$\varepsilon^{\ast}_{1}$ with respect to fragment $A_{1}$ is related
to the intrinsic excitation energy of the composite system and the
driving potential of the DNS. The intrinsic excitation energy is
provided by the kinetic energy of the relative motion, in which the
barrier distribution of colliding system is introduced in the
dissipation process. The motion of nucleons in the interacting
potential is governed by the single-particle
Hamiltonian.$^{\cite{Fe2}}$ Quasi-fission rate $\Lambda^{qf}$ and
fission rate (for heavy fragment) $\Lambda^{fis}$ are estimated with
one dimensional Kramers formula.$^{\cite{Ad2,Gra}}$ After reaching
the interaction time in the evolution of
$P(A_{1},\varepsilon^{\ast}_{1},t)$, the formation probability at
Coulomb barrier $B$ and angular momentum $J$ is given by
\begin{equation}
P_{CN}(E_{c.m.},J,B)=\sum_{A_{1}=1}^{A_{BG}}P(A_{1},\varepsilon^{\ast}_{1},\tau
_{int}(E_{c.m.},J,B)),
\end{equation}
where the interaction time $\tau _{int}$ is obtained by deflection
function method.$^{\cite{Li2}}$ So we can obtain fusion probability
as
\begin{equation}
P_{CN}(E_{c.m.},J)=\int f(B)P_{CN}(E_{c.m.},J,B)dB.
\end{equation}
For the barrier distribution function $f(B)$, we take an asymmetric
Gaussian form in the calculation. The survival probability of the
thermal compound nucleus is given by
\begin{equation}
W_{sur}(E_{CN}^{\ast},x,J)=P(E_{CN}^{\ast},x,J)\prod\limits_{i}\left(
\frac{\Gamma _{n}(E_{i}^{\ast},J)}{\Gamma
_{n}(E_{i}^{\ast},J)+\Gamma _{f}(E_{i}^{\ast},J)}\right) _{i},
\end{equation}
where the $E_{CN}^{\ast},J$ are the excitation energy and spin of
the compound nucleus, respectively. The realization probability
$P(E_{CN}^{\ast },x,J)$, neutron evaporation width $\Gamma _{n}$ and
fission width $\Gamma _{f}$ are calculated with the statistical
evaporation theory.$^{\cite{Fe2}}$

The partial cross sections of the capture, fusion and evaporation
residue are written respectively as
\begin{eqnarray}
&\sigma_{cap}(E_{c.m.},J)=\frac{\pi \hbar^{2}}{2\mu
E_{c.m.}}(2J+1)T(E_{c.m.},J), \\
&\sigma_{fus}(E_{c.m.},J)=\frac{\pi \hbar^{2}}{2\mu
E_{c.m.}}(2J+1)T(E_{c.m.},J)P_{CN}(E_{c.m.},J), \\
&\sigma_{evr}(E_{c.m.},J)=\frac{\pi \hbar^{2}}{2\mu
E_{c.m.}}(2J+1)T(E_{c.m.},J)P_{CN}(E_{c.m.},J)W_{sur}(E_{c.m.},J).
\end{eqnarray}
Within the framework of the DNS model, we have analyzed the partial
cross sections in the three stages for the reaction
$^{48}Ca+^{248}Cm$ at incident c.m. energies 201.91 MeV (solid line)
and 211.91 MeV (dashed line) as shown in Fig.1 (a)-(c),
corresponding to excitation energies of compound nucleus 35 MeV and
45 MeV respectively with the relation $E^{\ast}_{CN}=E_{c.m.}+Q$.
The reaction $Q$ value is given by $Q=\Delta M_{P}+\Delta
M_{T}-\Delta M_{C}$, and the corresponding mass defects are taken
from Ref.$\cite{Mol}$ for projectile, target and compound nucleus,
respectively. We can see that the dependence of the partial cross
sections on angular momentum is quite different. At the considered
excitation energies, higher angular momentum has larger contribution
for capture partial cross section. However, the maximal positions of
the fusion and evaporation residue partial cross sections begin to
remove towards lower angular momentum since fusion probability and
survival probability decrease rapidly with increasing angular
momentum. In Fig.1 (d) we take a sum of the evaporation residue
partial cross section over maximal angular momentum until
$J_{max}=30$ in 2n-4n evaporation channels. Calculated results can
reproduce experimental data except for the 4n evaporation channel
which predicts too high excitation energy. It may be related to the
incident energy loss in the target or to the calculated $Q$ value.
As a whole, the maximal values of the calculated $\sigma_{ER}$ are
consistent with the experimental results.$^{\cite{Og2}}$


With the same method as in Fig.1 (d), we have calculated the
evaporation residue excitation functions of the cold fusion reaction
$^{82}Se+^{209}Bi$ and the reaction $^{48}Ca+^{247}Bk$ in 2n-4n
channels to synthesize superheavy element 117 as shown in Fig.2. The
maximal production cross sections are 0.015 pb and 0.86 pb for the
reactions $^{82}Se+^{209}Bi$ in 1n channel and $^{48}Ca+^{247}Bk$ in
3n channel, corresponding to excitation energies 12 MeV and 37 MeV,
respectively. 2n channel and 4n channel have smaller production
cross sections, which are 0.61 pb and 0.48 pb, respectively,
corresponding to excitation energies 26 MeV and 47 MeV. Therefore,
it is favorable to synthesize new superheavy element 117 in 3n
evaporation channel for the systems $^{48}Ca+^{247}Bk$. Isotopic
trends to produce superheavy element 117 in cold fusion reactions
based $^{209}Bi$ target and in $^{48}Ca$ induced reactions are
analyzed using the DNS model in Fig.3. It is shown that isotope
projectiles $^{77}Se$, $^{79}Se$ and isotope targets $^{248}Bk$,
$^{249}Bk$ have larger production cross sections. $^{48}Ca$ induced
reactions have larger production cross sections than cold fusion
reactions. Here we only pay attention to 3n evaporation channel
since it has larger cross section than other evaporation channel.
Isotopic dependence of production cross section are mainly affected
by driving potential, neutron separation energy and shell correction
energy. Usually the incident channel with more asymmetric
combination (larger mass asymmetry degree of freedom) has larger
fusion cross section due to smaller inner fusion barrier. Smaller
neutron separation energy and higher fission barrier (mainly coming
from shell correction for superheavy nucleus) will increase the
survival probability of the formed compound nucleus. The reaction
$^{80}Se+^{209}Bi$ forms the element 117 with 172 neutrons, which is
a sub-magic number. In the case the one neutron separation energy is
larger, which leads to smaller survival probability, so as to
smaller evaporation residue cross section. Neutron-rich combinations
will reduce neutron separation energy, but it does not guarantee the
fusion probability to be larger. There exists the competition
between $P_{CN}$ and $W_{sur}$ as a function of the neutron richness
of the system. Pairing effect also affects the neutron evaporation
width and the fission width. It may be good way to synthesize
superheavy nucleus using radioactive nuclear beam in the future due
to larger survival probability of formed compound nucleus.


Superheavy elements have been successfully synthesized using
fusion-evaporation reaction in laboratories in the world. Usually
double magic or nearly double magic nuclei are selected to
synthesize superheavy nucleus because these combined systems have
larger $Q$ value, which can reduce excitation energy of compound
nucleus so that fusion probability and survival probability
increase, such as $^{48}Ca$ induced reactions in Dubna, and cold
fusion reactions based on $^{208}Pb$ or $^{209}Bi$ target at GSI. We
calculated other reaction systems to produce superheavy element 117
as shown in Fig.4, such as (a) $^{45}Sc+^{246}Cm$(solid line) and
$^{45}Sc+^{248}Cm$(dashed line), as well as (b) $^{51}V+^{244}Pu$,
(c) $^{55}Mn+^{238}U$ and (d) $^{59}Co+^{232}Th$ in 2n-4n channels.
These nuclei have longer half-lives and are possible to use in
laboratory with intense beams. Calculations give that the reactions
$^{45}Sc+^{246}Cm$ and $^{45}Sc+^{248}Cm$ have larger production
cross sections in 3n and 4n evaporation channels than other systems
since lower inner fusion barriers can increase the fusion
probability, which are suitable to synthesize the new element.
However, the target material $^{246,248}Cm$ are very difficult to be
prepared. To prepare $^{244}Pu$ is relatively easy, but one has to
face the smaller production cross section of 0.6 pb. In Table 1 we
list optimal excitation energies, maximal evaporation residue cross
sections as well as transmission, fusion and survival probabilities
at $J=0$ in 1n-4n evaporation channels for all of the possible
combinations to produce the element 117.


In summary, possible combinations are analyzed to synthesize new
superheavy element 117 within the framework of the DNS model. The
partial cross sections in the capture, fusion and evaporation
process for the reaction $^{48}Ca+^{248}Cm$ are calculated as a test
of the model. Within error bars experimental data can be reproduced
very well. Isotopic trends of the production of element 117 in cold
fusion reactions based $^{209}Bi$ target and in $^{48}Ca$ induced
reactions are investigated systematically. Calculated results show
that the reactions $^{48}Ca+^{248,249}Bk$ in 3n evaporation channels
and $^{45}Sc+^{246,248}Cm$ in 3n and 4n channels are favorable to
produce the element. The corresponding excitation energies are also
given in the letter.

\newpage

\begin{figure}
\begin{center}
{\includegraphics*[width=0.8\textwidth]{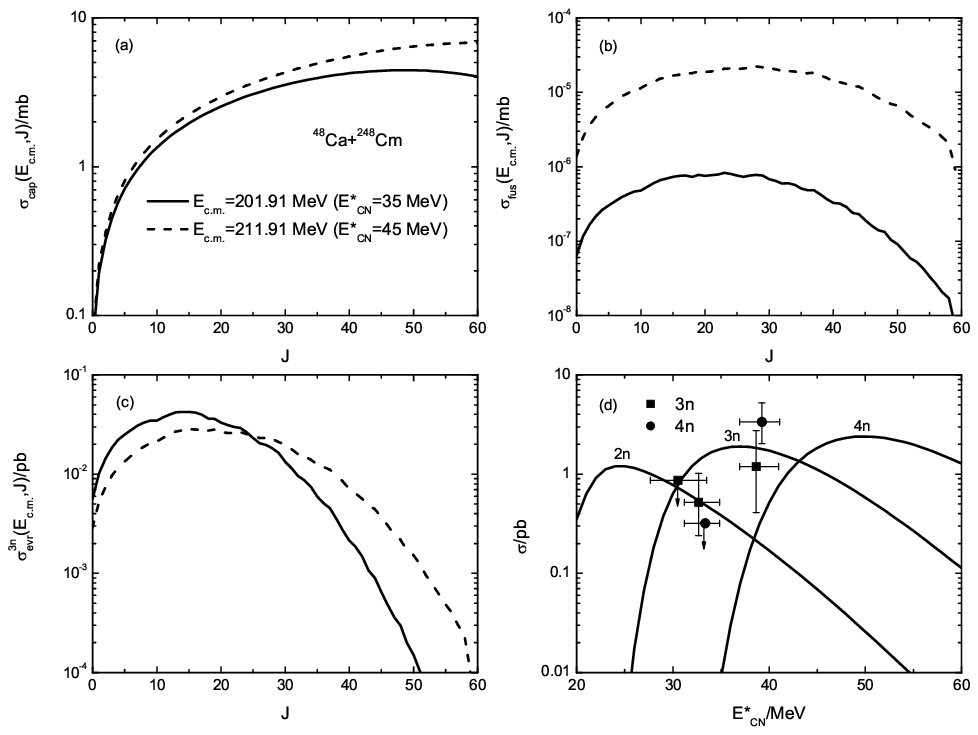}}
\end{center}
\caption{Angular momentum dependence of the partial cross section in
the process of capture (a), fusion (b) and evaporation residue in 3n
channel (c) for the system $^{48}Ca+^{248}Cm$ at incident c.m.
energies 201.91 MeV and 211.91 MeV respectively, and evaporation
residue excitation functions in 2n-4n channels compared with
experimental data (d) $^{\cite{Og2}}$.}
\end{figure}

\begin{figure}
\begin{center}
{\includegraphics*[width=0.8\textwidth]{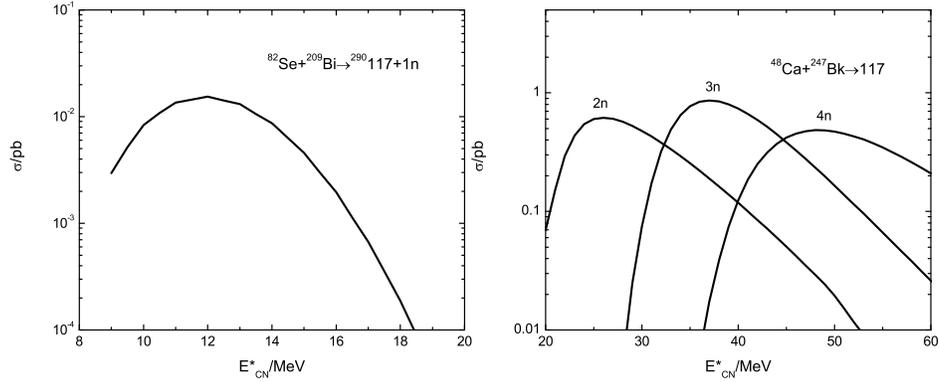}}
\end{center}
\caption{Evaporation residue excitation functions of the cold fusion
$^{82}Se+^{209}Bi$ and the reaction $^{48}Ca+^{247}Bk$ in 2n-4n
channels to synthesize superheavy element 117.}
\end{figure}

\begin{figure}
\begin{center}
{\includegraphics*[width=0.8\textwidth]{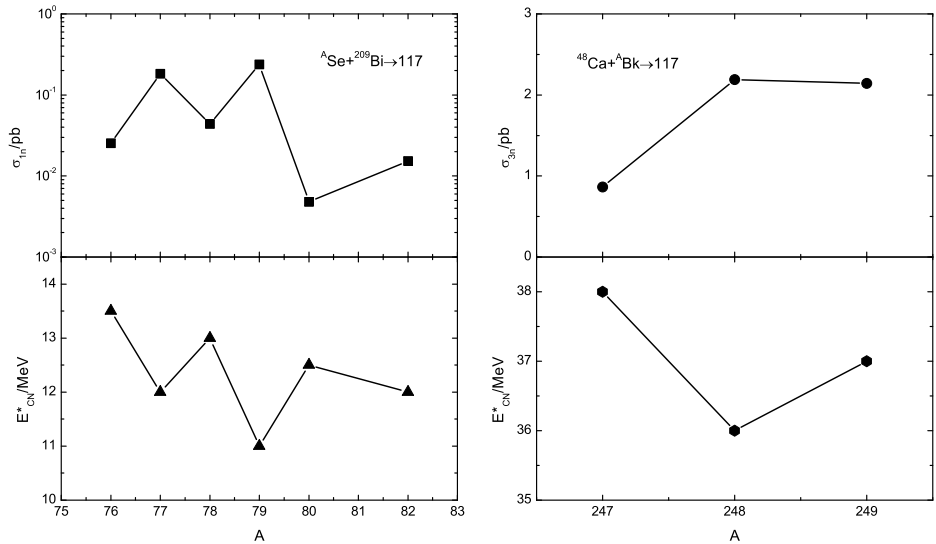}}
\end{center}
\caption{Isotopic dependence of maximal production cross sections in
1n and 3n channels (upper panel) and corresponding optimal
excitation energies (lower panel) to synthesize superheavy element
117 in cold fusion reactions based $^{209}Bi$ and in $^{48}Ca$
induced reactions.}
\end{figure}

\begin{figure}
\begin{center}
{\includegraphics*[width=0.8\textwidth]{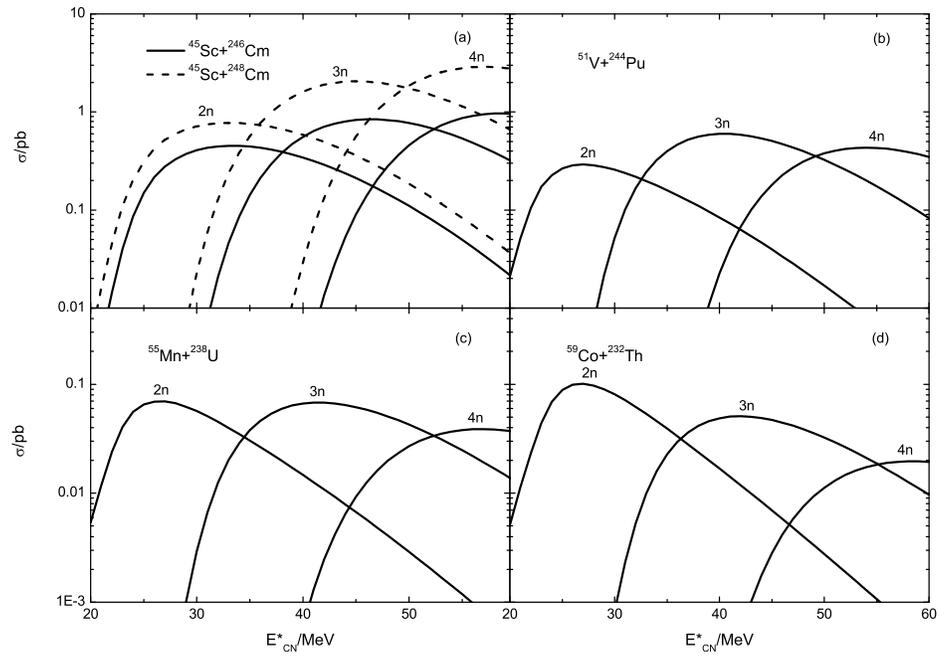}}
\end{center}
\caption{Comparisons of calculated evaporation residue excitation
functions in 2n-4n channels for the reaction systems
$^{45}Sc+^{246}Cm$(solid line) and $^{45}Sc+^{248}Cm$(dashed line),
$^{51}V+^{244}Pu$, $^{55}Mn+^{238}U$ and $^{59}Co+^{232}Th$ to
produce superheavy element 117.}
\end{figure}

\begin{center}
\begin{table}
\caption{Comparisons of calculated optimal excitation energies,
maximal evaporation residue cross sections and transmission, fusion,
survival probabilities at $J=0$ in 1n-4n evaporation channels.}
\begin{tabular}{|c|c|c|c|c|c|c|}
\hline
Reactions &Channels &$E^{\ast}_{CN}$/MeV &$T(J=0)$ &$P_{fus}(J=0)$ &$W_{sur}(J=0)$ &$\sigma_{ER}$/pb \\
\hline
$^{48}Ca+^{247}Bk$ &2n &26 &0.62 &7.19$\times10^{-9}$ &2.09$\times10^{-3}$ &0.61 \\
$^{48}Ca+^{247}Bk$ &3n &37 &0.88 &5.95$\times10^{-7}$ &1.51$\times10^{-5}$ &0.86 \\
$^{48}Ca+^{247}Bk$ &4n &47 &0.97 &9.69$\times10^{-6}$ &7.00$\times10^{-7}$ &0.48 \\
$^{48}Ca+^{248}Bk$ &3n &36 &0.86 &1.47$\times10^{-6}$ &5.13$\times10^{-5}$ &2.19 \\
$^{48}Ca+^{249}Bk$ &3n &36 &0.85 &1.24$\times10^{-6}$ &3.54$\times10^{-5}$ &2.14 \\
$^{45}Sc+^{246}Cm$ &2n &33 &0.45 &1.07$\times10^{-6}$ &3.43$\times10^{-5}$ &0.45 \\
$^{45}Sc+^{246}Cm$ &3n &46 &0.81 &7.06$\times10^{-5}$ &4.77$\times10^{-7}$ &0.84 \\
$^{45}Sc+^{246}Cm$ &4n &58 &0.96 &6.91$\times10^{-4}$ &4.31$\times10^{-8}$ &0.96 \\
$^{45}Sc+^{248}Cm$ &2n &33 &0.39 &2.24$\times10^{-6}$ &4.07$\times10^{-5}$ &0.77 \\
$^{45}Sc+^{248}Cm$ &3n &45 &0.74 &1.23$\times10^{-4}$ &8.95$\times10^{-7}$ &2.06 \\
$^{45}Sc+^{248}Cm$ &4n &57 &0.94 &1.27$\times10^{-3}$ &7.99$\times10^{-8}$ &2.88 \\
$^{51}V+^{244}Pu$  &2n &27 &0.42 &2.41$\times10^{-8}$ &1.20$\times10^{-3}$ &0.29 \\
$^{51}V+^{244}Pu$  &3n &41 &0.79 &4.13$\times10^{-6}$ &4.13$\times10^{-6}$ &0.60 \\
$^{51}V+^{244}Pu$  &4n &54 &0.96 &7.44$\times10^{-5}$ &2.17$\times10^{-7}$ &0.43 \\
$^{55}Mn+^{238}U$  &2n &27 &0.48 &5.72$\times10^{-9}$ &1.03$\times10^{-3}$ &0.07 \\
$^{55}Mn+^{238}U$  &3n &41 &0.76 &6.39$\times10^{-7}$ &3.14$\times10^{-6}$ &0.068 \\
$^{55}Mn+^{238}U$  &4n &56 &0.94 &1.54$\times10^{-5}$ &9.58$\times10^{-8}$ &0.038 \\
$^{59}Co+^{232}Th$  &2n &27 &0.51 &1.47$\times10^{-8}$ &6.77$\times10^{-4}$ &0.101 \\
$^{59}Co+^{232}Th$  &3n &42 &0.80 &1.13$\times10^{-6}$ &1.49$\times10^{-6}$ &0.051 \\
$^{59}Co+^{232}Th$  &4n &58 &0.95 &2.07$\times10^{-5}$ &4.31$\times10^{-8}$ &0.019 \\
$^{76}Se+^{209}Bi$  &1n &13.5 &0.85 &1.54$\times10^{-9}$ &1.64$\times10^{-3}$ &0.025 \\
$^{77}Se+^{209}Bi$  &1n &12   &0.84 &1.55$\times10^{-9}$ &1.03$\times10^{-2}$ &0.18 \\
$^{78}Se+^{209}Bi$  &1n &13   &0.85 &9.26$\times10^{-10}$ &4.71$\times10^{-3}$ &0.044 \\
$^{79}Se+^{209}Bi$  &1n &11   &0.83 &5.31$\times10^{-10}$ &3.64$\times10^{-2}$ &0.24 \\
$^{80}Se+^{209}Bi$  &1n &12.5 &0.89 &2.08$\times10^{-11}$ &1.29$\times10^{-2}$ &0.0048 \\
$^{82}Se+^{209}Bi$  &1n &12   &0.91 &4.67$\times10^{-11}$ &3.14$\times10^{-2}$ &0.015 \\
\hline
\end{tabular}
\end{table}
\end{center}

\end{document}